\documentclass[12pt]{article}

\usepackage{graphicx}
\usepackage{epsfig}
\usepackage{bm}% bold math
\usepackage{amsmath}
\usepackage{amsfonts}

\begin{document}

\title{OPERA Collaboration have observed phase speed of neutrino wave function, while advanced time displacement is mainly due to finite life time of pion}
\author{Shi-Yuan Li\\
School of Physics, Shandong University, Jinan, 250100, PRC}

\maketitle

\begin{abstract}
We show that the OPERA Collaboration have measured the phase velocity of the neutrino wave
function, based on the analysis of the experimental method.
%Since neutrino mass is very small, whether phase velocity or group velocity, if properly measured, should be quite near the light speed in vacuum.
On the other hand, the significant   advanced time displacement $\delta t$ reported by OPERA is mainly due to the finite life time  of the pions (w.r.t. its flying time  in the 1 km long tunnel) which decay to the muon neutrinos.
\end{abstract}

{\large \emph{1}}.
  The OPERA Collab. \cite{a2011zb} announced to have measured the `time of flight' T ($TOF_{\nu}$ in \cite{a2011zb}) of the neutrino from CERN to Gran Sasso.  Dividing the distance between these two places L (from BCT to OPERA \cite{a2011zb}) by   T, one can get a `velocity' whose value turned out to be larger than the speed of light in vacuum, c, i.e., L/T $>$ c.  Such a
result, at first glance, could be a challenge to Einstein's Special Relativity, though one is aware that c is one of the basic constant in  modern physics whose physical meaning is much more deeper than light's `speed' whose physical meaning is quite ambiguous.
 Only in the case of strict classical mechanics, when the position (element of some mathematical structure on which differential w.r.t. the parameter `time' can be defined) is considered as the basic observable of a particle, and as a single value function of time, the velocity is a well-defined quantity.
According to N\"{o}ther's  theorem, a basic and well defined  space-time related  observable is the generator of
some kind of space-time symmetry operation.   In  quantum theory, the movement of a particle is described by state vector, and  generally  position is not a good quantum number.
So in position representation we encounter a distribution described by wave function of the particle (leaving out the more complexity of particle creation and annihilation).  Hence velocity is too complex and case-dependent, generally  can not be taken as basic observable related to some kind of space-time symmetry, but as a quantity defined by others,
depending on the concrete cases.
E.g. (suppose  Lorentz invariance kept), for a free particle in energy-momentum eigenstate $|E~ {\bold P}>$, with $E^2=P^2+m^2$, its velocity can be defined by ${\bold P}/E$, but is not easy to relate to some unambiguous distance over time interval as in classical mechanics, even with the knowledge of the wave function.
 Another intuitive example is the electron bound in hydrogen atom, whose 3-momentum is  not a `good quantum number', i.e., not commuted with the Hamiltonian hence not the movement integral. In this case the above definition of velocity fails; and because  the electron is `off-shell', if we define some ratio between
  momentum and energy, or space and time, it is  very possible to get some `velocity' which is large than c.  One may call  the electron in such case as tachyon if one likes, but we
know that the basic physical principles like causality of the hydrogen system,  e.g., in process of transition between different energy levels, always keep. Hence we should pay attention  that  some `velocity' defined by  ratio between some measured space and time intervals is not always straightforward pointing to the basic principle about the property of space-time.
On aware of the above, we can well understand that even the terminology `speed of light in vacuum' is purely a traditional name. Since we should ask what kind of `light': a photon? a light pulse? etc.  For each case the definition of velocity varies. We know that c is further used, e.g., to scale the momentum, mass, energy of a free particle so that $E^2=P^2 +M^2$, etc., etc.
However, if the wave function of a particle can be deduced from some experiments, one of course can discuss some    `displacements' in space as well as time based on the wave function. Their ratio could be interesting provided
%since the complexity of `velocity' in Quantum theory framework, one needs to make
  a careful
investigation on what such kind of measurement  the experiment in fact makes, and what/whether basic physical principles can be further deduced from it.
%The latter is the main purpose this letter tries to pursue.
%In fact from our analysis we can see the L/T can not simply related to any physical %quantity.

 In the following we will recall that the phase speed of the wave function can very naturally be larger than c (section 2).  With the effort to analyze the key experiment method, i.e., fitting    the wave forms (average time distributions in the time interval of the extraction \cite{a2011zb}) of the proton as well as the neutrino event to get the time displacement w.r.t. `flying in the speed of c',  we show that the OPERA Collab. have measured the phase of the particle wave function and its  speed hence  is possibly  larger than c (section 3). Since neutrino mass is very small,    whether phase velocity or group velocity, if properly measured, both should not varying from c significantly, no matter larger or smaller than c.
  However,   due to   the finite life time of the pions (w.r.t. its flying time  in the 1 km long tunnel) which decay to the muon neutrinos, the wave form for neutrino is deformed from that of the proton. Such an effect is estimated to  lead to an advanced  time displacement
  of the order of 100ns, which coincide with the measured $\delta t $ reported in \cite{a2011zb}
  (section 4).

{\large \emph{2}}.
 The basic things measured by OPERA Collab. are two time distributions in two space-time  positions. %So first of all we show that they can not be simply related with the time of flight. On the other hand, in the framework of
%formal theory of scattering, esp. the static state scattering theory framework, we can well model this process and measurement.
%In such a model, one
At  position $x_0$ (CERN)
and time $t_0$, one  measures the proton to deduce  the wave function of the neutrino
 (for the problems see section 4) %at this space-time position
 and then at $t_1$ and  position
$x_1$ (Gran Sasso)  measures   %the %$\tau$ particle to get the value of
the wave function  by the OPERA detector %at that space-time point
(of curse we can only deduce the wave function by measuring its mudulus square).
To make the discussions simple, we do not discuss the neutrino oscillation.
The good time correspondence of the neutrino events and the proton extractions as presented by the Collab. \cite{a2011zb} indicates that the neutrino is well propagating and could be
modeled as a plane wave for the simplest consideration.
We will employ wave packet to describe the neutrino in Section 3 and
show  that the measured result L/T is possibly understood as  the phase speed of the propagating wave function since their key measurement is to calibrate the pace/phase for the particle in these two space-time positions.  L/T  larger than c is
the straightforward result from that the physical velocity of neutrino is smaller than c and well obeys the energy-momentum relation of a massive particle. Suppose a likely experiment is done for a much more massive   `dark matter' particle, such a phase speed may be significantly larger than c.
%In principle  can be considered as the phase speed of the 'matter wave', which is necessarily not smaller than c.
However, this does not to say such measurement is meaningless. Detailed analysis on such measurement may help to improve the experiments on neutrino oscillation or even absolute value of neutrino mass.

%In principle this measurement is the  measurement  of the phase displacement.
%So first we describe the neutrino employing exactly a plane wave as its wave function %(of the exact quantum mechanics meaning)
 %to demonstrate the above facts. %This discussion can be tracked back to de Broglie.....

 %So here we can consider %the two measurements as measure the plane wave in two places and times.
 %Because the measurements show that the plane wave is well propagating, %, we can deduce from %these two measurement (we see they show that the probability distribution of CERN and Gran %Sasso at thees two times is the same).
 %We can easily write
 The wave function is
\begin{equation}
\label{WF}
e^{-i\omega t + i \bold{\kappa \cdot x}},
\end{equation}
and we find that
\begin{equation}
\label{PV}
\frac{\omega}{|{\bold\kappa}|}=\frac{|\Delta {\bold x}|}{\Delta t}=\frac{| {\bold x_1-\bold x_0}|}{ t_1-t_0}
\end{equation}
This is the well-known  phase velocity of the wave of eq. (\ref{WF}).
Here one  encounters the similar problem that Louis de Broglie encountered
almost 100 years ago  \cite{deb}. %We know that when we
%input our knowledge on the particle to
When $\omega$ and $\kappa$ correspond to energy and momentum of a particle,  according to the Einstein relation, we
will find that this phase velocity is never smaller that c but always larger, except for light it  equals to c.
And the wise solution of employing more reasonable group velocity by de Broglie is the key corner stone for his matter wave
hypothesis  which is one of the most important turning points from Bohr's old quantum  theory to the modern  quantum mechanics.
% and gains infinite awards for him like the Nobel prize and le Grand Croix de  Ordre national %de la Legion d'honneur.

{\large \emph{3}}.
 A plane wave as above
% we  only measure a flat PDF which is dfferent from \cite{a2011zb}). And the plane wave
% the above is completely wrong as you will clarify that the timestructure of PDF is not tahty of the WF but just
%numbers which turns out can mark the phased!
  extends infinitely. If nothing can mark its phase,  our discussions
   in Section 2 are  just for an `ideal experiment'.
   By the above we only show that  for a plane wave, the interval between two space-time positions with the same phase could be space like.
   For the mechanical wave like that on a long shaking rope, we can see the oscillation (say, up and down) of the points of the rope  with different phases, so that we can observe the propagation. It turns out that the  time structure of the proton
 PDF (probability density function \cite{a2011zb}, but in fact  number density as to be clarified) and that of the OPERA neutrino events  can mark the phase  to be measurable, which we will make clear in the following based on the analysis of the experiment.
 It  is our key point in this section to make clear what the experiment measures with a simple model.
 Now it is more of reality to describe the propagation of neutrino by wave packet,
\begin{equation}
\label{WPWF}
\Psi (x, t)=\int  dk~\Psi(k)~  e^{-i \omega t + i k x}  ,
\end{equation}
where $\omega=\omega(k)$  is defined  by the on-shell Einstein relation for four momentum
(In this section the space related variables like $k$, $x$ are not specially denoted to show
they are SO(3) vectors but indicated).
Here we do not  discuss the details of $\Psi (k)$, only
this wave packet is assumed well localized: For space, even the  whole neutrino source system of CERN at
Prevessin and the detectors in laboratory at Gran Sasso
can be taken as point comparing to the distance between these two places.
 So here we need not discuss how well the position localized which is in fact well investigated by OPERA Collab. For time, we have no problem to say that  this particle is localized much smaller than $10.5 \mu s$, based on  the PDF time structure.
 Same as \cite{a2011zb}, we employ the time when the proton going through the BCT to mark the time of the wave packet (e.g., the central peak of the wave packet of Eq. (\ref{WPWF})).
  Taking into account  the  reaction time of the  BCT or
   %and the well corresponding $\tau$ signal which is the key measurement of the experiment
   by carefully analyzing the PDF, one can extract a basic time length $l_T$
   (say, around 5 ns according to Fig.4 OF \cite{a2011zb}). Here a proper $\Psi (k)$ need to lead that  $\Psi (x, t+\Delta t)$ vanishing when $|\Delta t| > l_T$.
   Since the wave packet could expand en route of propagation, so this requirement is only
   for the time and place at CERN, where the  protons  employed to produce neutrino by its reaction with the target is carefully measured, one of the most important instrument for the this measurement  is the BCT \cite{a2011zb}.

This is the key point here. We would like to remind the reader that if he/she understood that we assign the distribution
of the PDF or such kinds as  the time distribution of the  wave (packet) function of each proton as well as that of the neutrino, he/she completely misunderstood.
Particles produced in CERN in the processes described in Section 2 of \cite{a2011zb}, of which proton is the most
 especially  typical, can all be considered as that their wave packet size is as small as
possible (in time and space)  so that  the inner structure
 of the wave packet function is not possible seen by all detectors. The PDF time structures  exactly is the numbers of  particles  varying  with time.
However, this varying value can mark
%timestructure of PDF is not tahty of the WF but just
 %this different numbers which turns out can mark
 the PHASE of the wave packet of the particle.

Since in no way to assume any difference among all  the protons,
we can employ the same form of wave packet to describe all the protons as well as neutrinos.
Only that, each of the particles (proton/neutrino) corresponding to each `point'
 ($l_T$ is the unit) t' at the time axis of PDF distribution    % is one wave packet for the
    has a time displacement w.r.t. the beginning of the extraction:
\begin{equation}
\label{PWF}
e^{-iHt'} \Psi(x,t)= \int dk ~  \Psi(k) ~ e^{-i \omega (t+t') + i k x} ,
\end{equation}
 with $0<t'<10.5 \mu s$.
 This phase in other word can be called the pace of the protons/neutrinos, i.e.,
 to mark their order of `step' in time of  reactionreaction and reaction of each extraction.
 (Actually this only marks the neutrino's mother particles, mainly pions and kaons, the problems caused by the infinite life time during the decay is discussed in details in section 4.)
 %We here have,
  As the experiment has put together and makes statistics for all the particles in different `extractions',  they are to project on a same t' region by
 plus/minus n time of 50 ms (the time interval between extractions) for their real time t.  Here we emphasize again  that the wave packet is so well located
 that any `interference' between different wave packets (particles) are neglected, as can be assumed by the experiment (or effects   as beam interaction has been corrected by experiment).
 We  emphasize this is to point out that those kind of effects are of NO relation with our discussions.   And we also point out that though in experiment each event and wave form has been stamped by the
 real time, but what to be used in analysis is the averaged results obtained by large number of samples of extractions and events projected together on the interval of the extraction. We in fact can not say which proton (neutrino) has which phase but just employ the different numbers of particles (events) in this duration to represent the fact that each wave packet of the particle has certain definite phase, marked by t', relative to the, e.g, the beginning time of the extraction.

The wave packet will expand during propagating, but the experiment show they still keep some
peak structure  so that can be  detected in Gran Sasso at some certain time. This exact
value of time  but  is only used to extract the t' to get the event  time structure  for comparing with the source time structure (PDF),
%The key measurement is to project all data also by plus/minus n time of 50 ms to get the %distribution within the $10.5 \mu s$
then  to calibrate the phase parameter corresponding to t'.  The method is, as described by the OPERA Collab., to employ the maximum likehood method to coinside/tune the same pace for these two distributions.
This is just to determine which phase group of proton PDF the particles/events detected by OPERA belong to.  %which has set up in CERN.
Then it is possible  to find the same phase positions
 $(t_0, x_0)$ and $(t_1, x_1)$ when the phase/pace is
calibrated to coincide.
The space-time interval ratio hence can be
%These two measurements of distribution, when coincide each other after the fitting
%by the maximum likehood method can be
now  understood  as the velocity of the phase of the wave packet
marked by t' moves from
$(t_0,x_0)$  to $(t_1,x_1)$ (i.e., from CERN to Gran Sasso).
$x_1-x_0=L$ is the length of the baseline presented from \cite{a2011zb}, while $t_1-t_0=T$  is also possible deduced from the measured quantities in \cite{a2011zb}.
%The uncertainty of $t_0$, $t_1$ is $l_T$, while space can still be taken as point when
%size of the detector taken as the basic length.
In other words, the experiment is comparing the step order of the proton and reaction events,
so that to find two space-time position with the same phase.
Here we would like to emphasize again that the PDF and the  coinciding   neutrino  signal distribution is not the form of wave function but only show the fact that the wave packet's phase marked by t' properly propagates from $(x_0, t_0)$ to $(x_1, t_1)$.
Their interval could be space like as discussed in the following.

The movement of the wave packet from CERN to Gran Sasso,  %is written as
\begin{equation}
\label{WP}
e^{-iH(t_1-t_0)} \Psi (x_0, t_0, t')
\end{equation}
%%%%%%%%%%%%%%%%%%%%%%%%%%%%%%%%%%%%%%%%%%%%%%%%%%%%%%%%%
%Of course the measurement can only show their modulo squares are equal.
 %This experiment fact render to eq. (\ref{WP}).
%By inserting  eq. (\ref{WPWF}) to eq. (\ref{WP}), we can conclude based on
%$$\int \Psi(k) e^{-i \omega t_0 + k x_0}  dk=\int \Psi(k) e^{-i \omega t_1 + k x_1}  dk.$$
%In fact this do not rely on special $\Psi(k)$. But for a large set of functions which can
%make the wave packet well localized.
%So the conlcusion is the function in the exponential must equal and
%$$ L /T =(x_1-x_0)/(t_1-t_0)= \omega/k. $$
in practice is too complex to be employed to calculate the phase velocity and information (or `particle') velocity based on the real form of $\Psi (k)$, even we can get it. Here we only employ the (maybe too) simplified qualitative
expressions to discuss:
Suppose that the effective integral region of k (where $\Psi(k)$ significantly different from zero), $\Delta k$, is much smaller than k, i.e.,
\begin{equation}
\label{APCON}
\Delta k < < k,  ~which ~ also ~renders~
\Delta \omega < < \omega.
 \end{equation}
 This may be valid in  high energies.
 Now  we can  rewrite the wave function $\Psi (x, t, t')$ as
%\begin{equation}
%e^{-i (\omega+\Delta \omega) (t+t')+i (k+\Delta k)x} \int \Psi(k) dk.
%\end{equation}
%It is equals to
\begin{equation}
\label{BWF}
\Delta k \Psi(k) e^{-i \omega (t+t')+i kx}  e^{-i \Delta \omega t + i \Delta k x}.
\end{equation}
 $\Delta \omega  t'$ is higher order infinitesimal and neglected from Eq. \ref{BWF}, comparing to a long propagation time, since t' is fixed as a small value between $0~ to~ 10.5 \mu s$.
 So here t' only appear in the `carrier' wave.
From this quite simple expression we see the phase propagates with a velocity $\omega/k$, and
the same phase space-time position can be marked by t'. While the wave packet propagates at the group  velocity $\Delta \omega/ \Delta k$.  The above  simplification   is just a reproduction of the classical discussion from Brillouin' book, chapter 1 \cite{lb}.
In the most simple case the phase velocity is large than c while the group velocity is smaller, just to refer Section 2.

%With  the above simplification on single particle wave (packet) function we   mention more  %related with the experiment.
%It is easy to derive, with t
 The same narrow k distribution assumption (\ref{APCON}) for large k approximately leads to
\begin{equation}
\label{CLWF}
|\Psi(x, t+t')|^2 \cong \delta(v_g (t+t') -x),
\end{equation}
with $v_g=\Delta \omega/ \Delta k$ the group velocity in Eq. (\ref{BWF}).
This is just the distribution function of a classical particle and this  interpretes
why in most high energy experiments, one can treat the space-time movement of a single particle
 as classical with velocity
$P/E$, even with the fact it is almost in energy momentum eigenstate rather than position eigenstate.
%In the derivation, we need  the x and t  not very large (valid in CERN).
 Such a result also confirms what we have emphasized above, that in our treatment, we
need not to assume anything new from the conventional treatment about the proton beam,
as well as the pions, kaons, and neutrinos in the production tunnel  \cite{a2011zb}.
However, one should be aware that since in quantum mechanics,  a single microscopic particle is not able to be marked
and measured (e.g, the values of the position) many times without dramatically changing its state.  For a static
particle flow, this delta function description is of no difference from the plane wave description: no signal of  movement like  classical
  one  particle displacement in space-time can be observed.  %at the ideal case.

%%%%%%%%%%%%%%%%%%%%%%%%%%%%%%%%%%%%%%%%%%%%%%%%%%%%%%%%%%%%%%%%%%%%%%%%%%%%%%%%%%%%%%%

Now comes to the description of the time structure of the proton beam, which is also assumed
that of the neutrino. This is an inherent property of the proton beam system. Here we assumed that many times average (average particle number $<N(t')>$) can cancel the fluctuations from environment, i.e.,  $<N(t')>$ is just the PDF \cite{a2011zb}.
% and employ the   as normalization of the wave function to describe
So the particle system can be described as direct product of the  the single particle wave functions,
% is modified to be
with the interactions among them canceled.
\begin{equation}
\label{PWFS}
 \Phi(x,t+t')= \prod_{j=1} ^{<N(t')>} (\int dk_j ~  \Psi(k_j) )~ e^{-i \sum_{j=1} ^{<N(t')>}[\omega_j (t+t') + i k_j x]},
\end{equation}
  which renders
%  \begin{equation}
%\label{BWFS}
%\sqrt{<N(t')>} \Delta k \Psi(k) e^{-i \omega (t+t')+i kx}  e^{-i \Delta \omega t + i \Delta k x},
%\end{equation}
%and
\begin{equation}
\label{CLWFS}
|\Phi(x, t+t')|^2 \cong   \delta^{<N(t')>}(v_g (t+t') -x).
\end{equation}
Eq. (\ref{CLWFS}) shows that the BCT signal is proportional to the number of the protons $N(t')$ as they pass through the BCT at position x and time t+t'.
When we make the measurement for  the times structure of the number of particles to
 mark the same-phase space-time positions as the OPERA experiment, with the static particle flow rather than one particle movement, we are in touch with a
 space-time distribution B(x,t+t'), which is function of the phase, %taking into account
  i.e.,
 \begin{equation}
 B=B(\sum_{j=1} ^{<N(t')>}\omega_j (t+t') -  k_j x)
 \end{equation}
The experiments measured two space time positions with the same time structure, as one sees the different points of a rope with the same pace
of oscillation.

%%%%%%%%%Show a rope wave picture

The real information signal, is not the averaged number of $<N(t')>$, but a real time change
of this $<N(t')>$, which propagates with the velocity $v_g$.  Supposing  a distance large enough so that $x/v_g > > 50 \mu s$, then  at some time we change from the static flow from $<N_1(t')>$, to $<N_2(t')>$.
 and keep the new flow for a time $\tau$, with $50 < < \tau << x/v_g$, then change back to $<N_1(t')>$.
These two places can also communicate with light signal, and then comparing the propagation time for light and neutrino to get the velocity.  In fact this is just the
 analysis based on supernova 1987A observation \cite{sup87a}.

%At the same time, we find from Eq. (\ref{PWFS}) that

%That is all need to write down.

Such a simple model shows what the measurement the  OPERA Collab. have got.
%In one sentence: It is NOT strange if this  $\frac{|\Delta {\bold x}|}{\Delta t}$ larger than %c but could be strange if it is smaller.
Of course, further and deeper analysis, e.g., considering the propagation as pulse and employ more practical formulations, especially combined with oscillation, could help us to improve the measurement of the neutrino sector.
%Here we would like to remind the community that
%the complexity of wave speed. A complex model is easy to lead some measured velocity larger %than c \cite{}

The above discussion in fact is not restricted to the neutrino sector: It is a very general discussion based on quantum mechanics principles. However, other particle is not possible interact so weak so that can not travel such a long distance.
On the other hand, it is not very easy to discuss the causality  from the formulae here, since the causal or Feynman propagator is composed of positive and negative frequency parts, or say retarded as well as advanced parts to keep causality. Here we only employ positive frequency part. So a more practical formulation to describe the wave packet of a relativistic
particle in the quantum field theory framework still need exploration. And a precise observable in quantum field theory
 to define the pace of the particle is also in need.

{\large \emph{4}}.
 Since the neutrino mass is very small, whether the `velocity' measured by OPERA is the
phase one or group one, it should be very near the value of c, not as significantly
different as reported in \cite{a2011zb}.  Once the physical meaning of the velocity measured
by OPERA is clear, the question on what systematics which is missed by the OPERA data analysis cause the advanced time displacement
$\delta t \sim 60 ns$ need to be clarified.
Such an investigation is very important for this kind of measurements is reported to be
projected by other Collab., such as MINOS.
Here we show that the systematics comes from the finite
life  of the pions which decay to the muon neutrinos, $\nu_\mu$'s.

As is introduced by \cite{a2011zb}, the $\nu_\mu$ is produced in a 1 km long tunnel, from the decay of  the production of the proton-target collisions, mainly pion and kaon. In hadron-hadron collisions,  pion production is dominant. However, for higher energy region, the rate of kaon increases.  For a neutrino with certain energy,  the energy of its mother pion or kaon  is not quite different in average, but the $\gamma=E/m$ for kaon is much smaller than that of pion.
This leads to that the effective (Lorentz retarded) life time of the kaon is much smaller than that of the pion and two results are deduced: 1) This effectively increase the contribution of kaon to $\nu_\mu$; 2) the kaon can be taken as instantly decayed so that the $\nu_\mu$ from kaon can be considered as having the same wave form as the proton.

Let's give the numerical estimation for the $\gamma$'s and hence the life times of various particles which
decay to $\nu_\mu$ with energy 17 GeV as example, comparing with their flying time
in the 1km tunnel, $T_f \sim 0.3 \times 10^{-5}$.
For the case of pion, dominantly decaying to $\mu \nu_\mu$,  we
choose the case that the momentum of  $\nu_\mu$ in the same direction of  that of the pion.
We find that $\gamma$ must larger than several hundreds and its life time $T_\pi$ is
larger than  $10^{-5} s$,
which is comparable  with (and larger than) $T_f$, hence the neutrinos from pions have a  deformed  distribution from that of the proton.
For the Kaon sector,
with the same example case and same method, we can get the conclusion $T_K \sim 10^{-7} s < <T_f$
and hence the above conclusions.
But the $K_s$ should be treated separately. Since it mainly decays into two pions
instantly (comparing to $T_f$), so its case is same as that of the pion.

%with a much larger mass hence much smaller $\gamma$,
%its life comparing to the typical time interval of the extraction as well
%as the time of fly in the tunnel, can be taken as instant decay, up to
%
%several per cent corrections. This sector so can be considered to has the
%
%same wave form as that of the proton.

Other particles can also contribute, of them muon is very abundant.
For the case of muon, it decays into $\nu_\mu$ through a 3-body channel.
Take the case that $\nu_\mu$ energy is  half of the mass of muon and
momentum is parallel to the muon momentum, one can get the $\gamma$ is of
the order $10^2$. So $T_\mu$ is of the order $10^{-4}s   > > T_f$.
So for simplicity and a coarse estimation we can ignore the contribution
from the muon. If taken into account, the muon number can be taken as
constant during $T_f$ so will not lead to deformation of the wave form.

So the time of fly in the tunnel $T_f$ is the very key point
leads to the systematics  for the  distribution.
Mainly the case of pion to $\nu_\mu$ has to be considered.
Since the neutrino production is proportional to the number of the pions,
one can propose a exponentially  decreasing distribution for the neutrino
production time in the tunnel.
Besides directly to calculate the wave form deformation caused by the exponential form
of the time distribution of the neutrinos from the pion, the effect can be estimated by
 the displacement of the average production time of these neutrinos
 from the flat ones in the
flying interval, with the latter $T_f/2$.
Only considering the effect of
pion, it leads to the value of the order of 100 ns, which is coincide
with the OPERA result. The result is calculated from the following equations:
\begin{equation}
\frac{\int^1_0 dt~ t~ e^{-t/T_{\pi}}}{\int^1_0 dt ~ e^{-t/T_{\pi}}}.
\end{equation}
Here for simplicity we take $T_f$ as unit.
We also emphasize that this effect is not sensitive to
he wave form of the proton,  whether the old long time one and the new short time (pulse) one.
%In fact, we also point out that the time interval of 500ns only
%corresponding to 150m flying distance, so several pulses are mixed in the
%tunnel!

However, here we must point out that such an effect is significant but missed by the Collab. CRUCIALLY depends on the special
property of this kind of measurement and even more CRUCIALLY depends on the understanding for the physical meaning of this experiment: First, as clarified in section 3,  this experiment compares phase/pase (`wave form') to measure the
phase velocity. Second, the recorded neutrino events are quite rare, statistically the number of neutrino events recorded by OPERA per proton extraction is much much smaller than one. In such a special case, the decay process in the tunnel for each extraction is not possible to be tracked
by the neutrino events (requiring number of neutrino events per extraction $\geq  2$), so statistically the phase displacement in the whole  tunnel
flying will deform the recorded neutrino event wave form. Needless to say,
if one insists to understand this experiment as to directly measure the flying of each neutrino recorded
by OPERA, in the
same way for  a classical one, the decay effect is very
small because $1-\beta_{\pi} \sim 10^{-4}$ and pion only fly not more than 1km, which  HAS been taken into account in \cite{a2011zb}.

The MINOS \cite{minos} experiment is much less affected by the decay effect discussed above  since its energy  is much lower hence a much smaller $\gamma$. All the mother particles can be taken as to decay instantly.

{\large \emph{5}}.
 After the release of the OPERA data,  there have been more than 100 papers discussing this result \cite{ar}.
This has stimulated the study/review  of the Lorentz invariance violation in various ways, which could be a good
window for the future new physics.
 On the other hand, there are several papers devoted to study the systematics of the experiment or suggesting the velocity could be
 unphysical one. However, such kind of investigation  should include several key points: first to clarify what the observable is measured
 by analyzing the basic experimental method; second  to see whether the measured value agree or not with the suggested, if not, where is the
 systematic.  This paper is   trying to do in this way. This is a thing that one can not escape from  for exploring Lorentz invariance violation,
 not relating to any conservative attitude of believing Einstein, but  something learned from Einstein's attitude to experiments, no matter sooner or later his theories are found to be broken through.

% however, it is not a good thing...
%Our paper aims to understand the experiments methods as well as its sytematyics,
%NO ANY relations about the beliefs, whether respecting  to Einstein or relativity
%...
%On the other hand, %without investigation on the measurement but
%thoughts like  Lorentz violation (other kinds of inertial frame transformation laws, caused e.g., from quantum effects of gravity, etc.),
%or another Lorentz transformation parameter larger than c existing only for the neutrino sector,  etc.,  are of course valuable. Papers on %likelytopics are bursting out superluminally at  almost 10/day in these several days, which are not possible to make a full citation here.

%However, based on the discussions above, here we want to call the attention on that the signals of these thoughts
%  calculated in the QUANTUM  field theory level,  especially related with collider
%physics and then a really solid relation with their thoughts as well as the experiment observable is set
%need to be further carefully investigated.
%One who interested in should pay more attention on this side, the author thinks this is what one should learn from Einstein, no matter sooner or later %his theories are found to be broken
%through one day.

\vskip 2cm
The author thanks  Prof. Dr. WANG Meng for explaining some aspects of the OPERA experiments.
This work is partially supported by NSFC(10935012), SFSD (JQ201101) and SDU (2010JQ006).

\end{document}